\documentclass[aps,amsmath,twocolumn,amssymb,floatfix,showpacs,superscriptaddress,footinbib]{revtex4-1}

\usepackage{graphicx}
\usepackage{dcolumn}
\usepackage{multirow}
\usepackage{booktabs}
\usepackage{bm,color}
\usepackage{braket}
\usepackage{amsmath,amssymb}
\usepackage[colorlinks,linkcolor=blue,hyperindex,CJKbookmarks]{hyperref}
\usepackage{epstopdf}

\def\be{\begin{equation}}
\def\ee{\end{equation}}

\def\nc2{\left( \begin{array}{c} n \\ 2 \end{array}\right)}

\def\beq{\begin{equation}}
\def\eeq{\end{equation}}

\def\R{{\vert{\bf R}_{i}-{\bf R}_{i+1}\vert}}

  \def\ket{\vert \vert  \{ \emptyset \} \rangle}
  \def\ket2{\vert \vert \otimes \{ R \} \rangle}

\def\.#1{\mathaccent 95#1}
\def\^#1{\mathaccent 94 #1}
\def\~#1{\mathaccent "7E #1}

\def\c#1{\mbox{\bf #1}}

  \def\ket{\vert \vert  \{ \emptyset \} \rangle}
  \def\ket2{\vert \vert \otimes \{ R \} \rangle}

\def\R{{\vert{\bf R}_{i}-{\bf R}_{i+1}\vert}}

  \def\ket{\vert \vert  \{ \emptyset \} \rangle}
  \def\ket2{\vert \vert \otimes \{ R \} \rangle}

  \def\ket{\vert \vert  \{ \emptyset \} \rangle}
  \def\ket2{\vert \vert \otimes \{ R \} \rangle}

\def\.#1{\mathaccent 95#1}
\def\^#1{\mathaccent 94 #1}
\def\~#1{\mathaccent "7E #1}

\def\c#1{\mbox{\bf #1}}

\def\be{\begin{equation}}
\def\ee{\end{equation}}

\def\nc2{\left( \begin{array}{c} n \\ 2 \end{array}\right)}

\def\beq{\begin{equation}}
\def\eeq{\end{equation}}

\def\R{\vec{R}}

\def\RR{\vec{R'}}

  \def\ket{\vert \vert  \{ \emptyset \} \rangle}
  \def\ket2{\vert \vert \otimes \{ R \} \rangle}

\def\.#1{\mathaccent 95#1}
\def\^#1{\mathaccent 94 #1}
\def\~#1{\mathaccent "7E #1}

\def\c#1{\mbox{\bf #1}}

  \def\ket{\vert \vert  \{ \emptyset \} \rangle}
  \def\ket2{\vert \vert \otimes \{ R \} \rangle}

  \def\ket{\vert \vert  \{ \emptyset \} \rangle}
  \def\ket2{\vert \vert \otimes \{ R \} \rangle}

  \def\ket{\vert \vert  \{ \emptyset \} \rangle}
  \def\ket2{\vert \vert \otimes \{ R \} \rangle}

\def\.#1{\mathaccent 95#1}
\def\^#1{\mathaccent 94 #1}
\def\~#1{\mathaccent "7E #1}

\def\c#1{\mbox{\bf #1}}

\def\be{\begin{equation}}
\def\ee{\end{equation}}
\def\etal{{\sl et.al}}

\begin{document}

\title{\bf{ Magnetization dynamics in Fe$_x$Co$_{1-x}$ in presence of chemical disorder}}

\author{Banasree Sadhukhan}
\email{banasree@kth.se}
\affiliation{ KTH Royal Institute of Technology, AlbaNova University Center, SE-10691 Stockholm, Sweden}
\affiliation{Department of Physics, Presidency University, 86/1 College Street, Kolkata 700073, India}
\author{Raghuveer Chimata}
\affiliation{Department of Physics and Astronomy, Uppsala University, Box 516, SE-75120 Uppsala, Sweden}
\author{Biplab Sanyal}
\email{biplab.sanyal@physics.uu.se}
\affiliation{Department of Physics and Astronomy, Uppsala University, Box 516, SE-75120 Uppsala, Sweden}
\author{Abhijit Mookerjee}
\email{The project was conceived by Late Prof. Abhijit Mookerjee.}
\affiliation{S.N. Bose National Centre for Basic Sciences, JD-III, Salt Lake, Kolkata 7000098,India}
\date{\today}

\begin{abstract}
In this paper, we present a theoretical formulation of magnetization dynamics in disordered binary alloys based on Kubo linear response theory interfaced with the combination of seamlessly three approaches; density functional based tight-binding linear muffin-tin orbitals,  generalized recursion and Augmented space formalism. We apply this method to study the magnetization dynamics in chemically disordered Fe$_x$Co$_{1-x}$ ($x$ = 0.2, 0.5, 0.8) alloys. We reported that the magnon energies decrease with an increase in Co concentration. Significant magnon softening has been observed in Fe$_{20}$Co$_{80}$ at the Brillouin zone boundary. The magnon-electron scattering increases with increasing Co content which in turn modifies the hybridization between the Fe and Co atoms. This reduces the exchange energy between the atoms and soften down the magnon energy. The lowest magnon lifetime in found in Fe$_{50}$Co$_{50}$, where disorder is maximum. This clearly indicates that  the damping of magnon energies in Fe$_x$Co$_{1-x}$ is governed by the hybridization between Fe and Co whereas the magnon lifetime is controlled by disorder configuration. Our atomistic spin dynamics simulations show a reasonable agreement with our theoretical approach in magnon dispersion for different alloy compositions.
\end{abstract}

\maketitle

\noindent
\section{Introduction.}
\par The dynamics and damping of magnetic excitations play a pivotal role in many modern day spintronic devices by exploring the nature of the magnon dispersion and spin transport in pristine metals, semiconductors and their alloys.   Band theory of ferromagnetism based on the itinerant electrons has successfully predicted the magnetic properties in metallic systems including transition metals, Heusler alloys, and rare-earth magnets.  But the theoretical investigations has been a controversial subject of modern science for the  dual character of the d-electrons.  Their ground state is described by the band-like itinerant electrons at $T=0$.  However,  it arises many open questions of more general finite temperature ab initio approach which include spin excitation  \cite{moriya2012electron,  moriya1980unified}. The spin wave theory of magnetism includes the fluctuations which switch to spin dynamics approach from band theory of ferromagnetism \cite{PhysRevB.54.1019,  PhysRevB.72.172410,  PhysRevB.99.104302,  eriksson2017atomistic, skubic2008method}.

First principles study of spin correlation function and magnon lifetime of disordered magnetic systems have become important because of both fundamental and technological interest. Recent advancement in experimental techniques have enabled us to probe the magnon dynamics \cite{dalosto,  hopster,  vollmer1,  vollmer2,  vollmer3,  zhang, zakeri,  antropov,  zakeri1}. In a recent experiment using spin-polarized electron energy loss spectroscopy (SPEELS) technique, it has been shown that magnon energies reduce (referred as magnon softening) for single ferromagnetic monolayer of Fe on W compared to the pure bulk Fe \cite{prokop}. The theoretical predictions, based on an itinerant electron model, are in contradiction with the above experimental finding. The theoretical limitations are in the lack of correctly understanding of the fundamental nature of spin-wave excitations. The structural relaxation between the Fe and W layers influences the hybridization between Fe and W states, which modifies the exchange interactions \cite{ander}. Disordered local moments (DLM) configuration reduces the exchange interaction strengths. This leads to significant magnon softening compared to bulk Fe. The magnetic ordering coming from Dzyaloshinskii-Moriya interactions (DMI) as a consequence of  spin orbit coupling also leads to magnon softening. For the bulk system, this effects can be ignored. 

Continuum and atomistic models have been developed to study the spin dynamics in real materials.  In micromagnetic simulations, a continuum model of magnetization is considered in a length scale much bigger than the interatomic distance.   \cite{fidler2000micromagnetic}. Regarding atomistic models based on first principles methods, one may directly solve time-dependent density functional equation, which requires a huge computational cost and time to make realistic prediction in materials.  The standard approach to simulate the time evolution of the spin texture is to propagate the Landau-Lifshitz-Gilbert (LLG) equation \cite{krieger2015laser, LLG}.  However, the Hamiltonian required for solving LLG equation contains several terms involving magnetic exchange and anisotropy. It should be mentioned that for a disordered magnetic alloy, the exchange interaction between atoms may depend strongly on the chemical composition \cite{shimizu1981itinerant,  PhysRevB.51.8958,  PhysRevLett.83.2062,  PhysRevLett.117.087205}. Therefore,  the magnon spectrum and lifetimes in a material depend on the configuration of the magnetic state via the local atomic environment. Another important parameter in LLG equation is the Gilbert damping parameter, which can be calculated by first principles electronic structure calculations. Two successful models in this regard are the breathing Fermi surface model (BFS)  \cite{BFS}  and the torque correlation model (TCM)  \cite{TCM}. Unfortunately, neither the models are parameter free, nor do they really shed light on the microscopic origins of damping. Finally, it was realized that damping arose from microscopic scattering processes \cite{brataas, liu, ebert}.  Ebert \etal\ calculated the Gilbert damping parameter for bcc Fe$_x$Co$_{1-x}$ alloy from CPA approach via linear response theory \cite{ebert}. But the microscopic origin of the damping is still unfolded. 

The dynamical relaxation of various complex systems have also been the focus of considerable interest. In particular, spin relaxation in a transition metal based alloys have received considerable attention over the years. The relaxation time for the ferromagnetic transition metals like Fe, Co and Ni, is strongly spin dependent. The spin-averaged relaxation time in those metals is much shorter than in the noble metals. The analysis of the peak-positions and broadening of magnon excitation provides us information on the magnon energy and lifetime, respectively.  Magnon energy decays exponentially like exp(-$\gamma$t/2$\hbar$) where $\gamma$ represents intrinsic line width of Lorentzian peak in magnon spectral density and $\hbar$ is the reduced Plank constant. The magnon lifetime $\tau = 2\hbar/\gamma$ is usually defined as the time in which the amplitude drops to  e$^{-1}$ of its initial value.  The is similar in concept of the quantum mechanical broadening parameter and lifetime of an energy eigenstate \cite{PhysRevB.80.241415,  PhysRevB.77.195412,  PhysRevB.47.9650,  sadhukhan2017disorder,  PhysRevMaterials.4.064602}.

\par The physical picture underlying the spin-dynamics method, which we present here,  is not  completely new,  but their implementations in presence of chemical as well as magnetic interactions within first principles accuracy are of fundamental importance.  The Augmented space formalism (ASF) has been used successfully to describe the effect of different disorder \cite{saha1996effects,  kaphle2015study,  PhysRevB.81.184205,  PhysRevB.79.224204,  PhysRevB.79.054201,  PhysRevB.75.134203,  PhysRevB.75.134202,  PhysRevB.50.13267,  PhysRevB.96.054203, sadhukhan2017,  sadhukhan2017effect}.  Here we apply ASF for the simulation of spin dynamics,  that enable the characterization of magnetic excitations in presence of chemical disorder within Kubo linear response formalism in combination with density functional based first-principles approach.  The Hamiltonian parameters obtained from first-principles theory provides a reliable tool for analysis and even prediction of complex collective modes of magnetic materials. Our aim is to probe the magnon dynamics in chemically disordered Fe$_x$Co$_{1-x}$ alloys.

\par The article is organized as follows : In Sec.~\ref{method},  we describe the Hamiltonian for spin dynamics for binary alloys in presence of chemical disorder,  using Augmented space recursion (ASR) approach to tackle the disordered Hamiltonian and techniques for the calculations of the adiabatic magnon,  lifetime,  and a atomistic scheme for numerical integration of the equations of motion.  We present the computational details to describe the dynamical spin response functions.  Sec.~\ref{result} contains the applicability of our method taking bcc Fe as a testing ground,  and results and discussions on Fe$_x$Co$_{1-x}$.  Finally,  we give conclusions and outlook in Sec.~\ref{conclusion} .

\section{Methodology} 
\label{method}     
\subsection{Spin transport at low temperatures}

\par  We analyse the spin transport in ferromagnetic alloys. We begin with a ferromagnetic sea as our unperturbed state. A spin flip is an excitation or fluctuation in that state. We use the XXZ Heisenberg Hamiltonian  with a random distribution of the exchange parameters in terms of creation and annihilation operators :
\begin{equation}
{ H} =  - \sum_{\R}\sum_{\RR} J(\R -\RR) \left(   a^\dagger_{\R}a_{\RR}+a^\dagger_{\RR}a_{\R}+\Delta\ \tilde{n}_{\R}\ \tilde{n}_{\RR}\right)
\end{equation}
where $\Delta$ is the anisotropic exchange parameter describing the magnon-magnon interaction, $J(\R-\RR)$ is the strength of the exchange interaction. $a^\dagger_{\R}$ , $a_{\R}$ are the magnon creation, annihilation operators respectively and $\tilde{n}_R = a^\dagger_{\R}a_{\R}$. The excitations in this model are described by spin waves or magnons that can be envisaged as spin-patterns against a uniform spin background moving on the underlying lattice.  Here we consider $\Delta \sim 0$ neglecting the magnon-magnon interaction within linear approximation.

\par The linear response to an external homogeneous disturbance is 
 described in terms of two particle Green's functions within the Kubo linear response theory \cite{kubo}.  If a spin system is disturbed by an external field, which causes a perturbation in XXZ Heisenberg Hamiltonian, then the spin response function $\Gamma (\R-\RR,t-t')$ is related to the disturbance by a spin-spin correlation function $C(\R-\RR,t-t')$ \cite{kubo, vis2, vis3} which is given by:
\begin{equation} 
\Gamma(\vec{R}-\RR,t-t') = \frac{i}{\hbar} \Theta(t-t') \langle\Phi_0\vert[{\sigma}(\vec{R},t),{\sigma}(\RR,t']\vert\Phi_0\rangle   \nonumber
\end{equation}
where $C(\R-\RR,t-t')=\langle\Phi_0\vert[{\sigma}(\vec{R},t),{\sigma}(\RR,t']\vert\Phi_0\rangle $,
${\sigma}(\R,t)$ is the spin operator, $\Theta$ is the Heaviside step function and $|\Phi_0>$ the ground state.  The dynamical structure factor $S(\vec{q},\omega)$ is the Laplace transform of the spin correlation function $C(\R-\RR,t-t')$.

\subsection{Recursive approach to dynamical spin response functions}

\par Our starting point is a dynamical variable described by the Hermitian operator $D(\vec{q}, t)$ following the time evolution such that $\vert \Phi (t)\rangle = D(t)|\Phi_0\rangle $.  Our guiding equation is the Kohn-Sham equation : 
\begin{equation}
 i\frac{\partial |\Phi(t)\rangle}{\partial t} =  H |\Phi(t)\rangle  \nonumber
\end{equation}
 We shall follow the recursion procedure described by Gagliano and Balserio \cite{gb} and,  Viswanathan and Muller \cite{vis1,  vis2, vis3,  vis4}. We first choose a
denumerable basis of representation $\{|\phi_n>\}$ and expand the ``wave function" in this basis :
\begin{equation}
|\Phi(t)\rangle=\sum_{n=1}^\infty\ D_n(\vec{q},t) |\phi_n\rangle \label{rec}  
\end{equation}

\par We begin with $|\phi_1\rangle=a^\dagger(\vec{q})|\Phi_0\rangle$  where $a^\dagger(\vec{q})=\frac{1}{\sqrt{N}}\sum_{\R} e^{-i\vec{q}\cdot\R} a^\dagger(\R)$.  Next step, $|\phi_2\rangle = H|\phi_1\rangle - \alpha_1|\phi_1\rangle $ and orthogonality leads to $ <\phi_1|\phi_2\rangle = 0\Rightarrow \alpha_1 =\frac{\displaystyle <\phi_1|H|\phi_1>}{\displaystyle <\phi_1|\phi_1>} $.  Finally for $n>2$
\begin{equation}
 |\phi_{n+1}> = H|\phi_n> - \alpha_n |\phi_n> - \beta_n^2|n-1>  \nonumber
\end{equation}
orthogonality  $<\phi_n|\phi_{n+1} > = 0 \Rightarrow  \alpha_n = \frac{\displaystyle<\phi_n|H|\phi_n>}{\displaystyle <\phi_n|\phi_n>}$ \quad  $\beta_n^2 = \frac{<\displaystyle \phi_{n+1}|\phi_{n+1}>}{\displaystyle <\phi_n|\phi_n>}$.  The $\{ \alpha_n,   \beta_n \}$ are the recursion parameters.

\par Substituting this in Eq.(\ref{rec}), we get :
\begin{equation} 
i \frac{\partial D_n(\vec{q},t)}{\partial t} =  D_{n-1}(\vec{q},t) - \alpha_n D_n(\vec{q},t) - \beta_{n+1}^2 D_{n+1}(\vec{q},t) \nonumber
\end{equation}
Taking the Laplace Transformation :
\begin{equation} 
(z-\alpha_n) {D}_n(\vec{q},z) - i \delta_{n0} = {D}_{n-1}(\vec{q},z) + \beta_{n+1}^2 {D}_{n+1}(\vec{q},z) \nonumber
\end{equation}
where,
\begin{equation}
{D}_0(\vec{q},z) = \frac{\displaystyle i}{\displaystyle z-\alpha_1 -\frac{\displaystyle \beta_1^2}{\displaystyle z-\alpha_2 - \frac{\displaystyle \beta_2^2}{
\displaystyle z-\alpha_3 \ldots}}}\label{a1} \nonumber
\end{equation}
The dynamical structure factor is then 
\begin{equation}
S(\vec{q},\omega)= \lim_{\delta\rightarrow 0} Re[{D}_0(\vec{q},\omega+i\delta)] \nonumber
\end{equation}
where $z=\omega+i\delta$.

\subsection{Dynamical spin response functions in the presence of disorder : the Augmented space approach}

\par Now, we introduce disorder in the exchange parameters $J(\R-\RR)$ by a set of random variables $\{n_{\R}\}$.  

\par  The randomness can be introduced in the exchange parameter through a local, binary random variable $n_{\R}$ ($\R$ can be occupied by  A or B type of atom for binary alloys) taking the values 0 and 1 with probabilities $x$ and $y=1-x$ respectively :
\begin{eqnarray*}
J(\vec{R}) = J^{AA}(\vec{R})n_{\R} n_{\R +\vec{x}}+ J^{BB}(\vec{R})(1-n_{\R})(1-n_{\R+\vec{x}})+ \\  J^{AB}(\vec{R}) \left(n_{\R}(1-n_{\R+\vec{x}})+(1-n_{\R}) n_{\R+\vec{x}}\right)
\end{eqnarray*}
where $\RR = \R + \vec{x}$

The Augmented space method \cite{11, 29, 30,31,32} replaces these random variables $\{n_{\R}\}$ by operators $\{N_{\R}\}$. The eigenvalues are the random numbers taken by the variables $\{n_{\R}\}$ and  the spectral functions are the probability densities of the operator  $\{N_{\R}\}$ \cite{config}.  The Augmented space is an outer product of real space $\Phi_{\vec{R}}$ (lattice space) and configuration space $\{\emptyset\}$ (random space).  Then,  the configuration (disorder) averaged spin response function is given by : 
\begin{eqnarray}
 \ll \Gamma({\vec{q}},z)\gg &  = &\langle \Phi_{\vec{q}} \otimes \{\emptyset\}\vert (z\tilde{I} -\tilde{H})^{-1}\vert
\Phi_{\vec{q}} \otimes \{\emptyset\} \rangle \nonumber
\end{eqnarray}
where $\tilde{H}$ is the disordered Hamiltonian.

 \par Now, the configuration averaged dynamical operator in disordered system can be expressed as a continued fraction of recursion coefficients $\{ \tilde \alpha_n,  \tilde \beta_n\}$ as :
\begin{eqnarray}
 \ll {D}_0(\vec{q},z)\gg  = \frac{\displaystyle i}{\displaystyle z-\tilde{\alpha}_1 -\frac{\displaystyle \tilde{\beta}_1^2}{\displaystyle z-\tilde{\alpha}_2 - \frac{\displaystyle \tilde{\beta}_2^2}{
\displaystyle z-\tilde{\alpha}_3 \ldots}}}   \nonumber
\label{a2}
\end{eqnarray}
Then, the configuration averaged dynamical structure factor is obtained by :
\be
 \ll  S(\vec{q},\omega) \gg= \lim_{\delta\rightarrow 0} Re  \ll {D}_0(\vec{q},\omega+i\delta) \gg  \nonumber
\label{sw}
\ee
The self-energy $\Sigma(\vec{q},z)$ via the recursion method suggested by Viswanath and M\"uller \cite{vis1,  vis2, vis3,  vis4} is given by: 
\be  
\Sigma(\vec{q},z) =  \frac{\displaystyle \tilde \beta_2^2}{\displaystyle z - \tilde \alpha_2 -\frac{\displaystyle \tilde \beta_3^2}{
\displaystyle z - \tilde \alpha_3 - \frac{ \tilde \beta_4^2}{\ddots z- \tilde \alpha_n- \tilde \beta_n^2 T(\vec{q},z)}}}  \nonumber
\ee
 The terminator $T(\vec{q},z)$ which reflects all the singularities in the response function,  is constructed from the calculated $\{\tilde \alpha_n,  \tilde  \beta_n \}$ for $n=1,\ldots N$ \cite{haydock1,  luchini,  beer,  magnus,  chahn,  vis4}.  Therefore, we used the square-root terminator terminator \cite{luchini,  beer} with the calculated recursion coefficients to taken into account the convergence of the continued fraction.  Im$[\Sigma(\vec{q},\omega)]$ counts the disorder induced broaden which gives the disorder induced lifetime $\tau$ of magnon state by Im$[\Sigma(\vec{q},\omega)]$ = $ 1/\tau(\vec{q})$

 \par We study the configuration averaged dynamical response function of chemically disordered binary alloys using with ASR within Kubo linear response theory \cite{11,29,30,31,32}.  ASR technique goes beyond the usual single site mean field, coherent potential approximation (CPA) like approach to study the environmental effects for chemically disordered alloys.  In particular,  the methodology we implemented here, is an admixture the tight binding linear muffin-tin orbitals (TB-LMTO) technique,  ASF  and generalized recursion (GR)  \cite{vis1,  vis2,  vis3, haydock}.  It provide an accurate computational framework for analysing  dynamical response properties.

 \par We have calculated configuration averaged dynamical structure factor $ \ll S(\vec{q},\omega) \gg $ of Fe$_x$Co$_{1-x}$ for three Co content ($x$ = 0.2, 0.5, 0.8).  This relates the response of the systems in presence of an external perturbation (here the chemical disorder) from the self energy.  We also calculated the magnon lifetime $\tau(\vec{q})$ of Fe$_x$Co$_{1-x}$. The Brillouin zone boundary in the [001] direction of FeCo is nearly 2 ${\AA}^{-1}$. We neglect the effect of magnon-magnon and magnon-stoner interaction. Here, we basically focus on the magnon-electron interaction to study the effects of chemical disorder on Fe$_x$Co$_{1-x}$.

\subsection{Atomistic spin dynamics}

\par To compare our results, we also calculated the magnon dispersion spectra of Fe$_{x}$Co$_{1-x}$ ($x$ = 0.2, 0.5, 0.8) from Uppsala Atomistic Spin Dynamics (UppASD) code \cite{asd1,asd2}.  The necessary exchange parameters for UppASD are calculated in the the framework of {\em ab-initio} density functional theory (DFT). They were done by means of the Korringa-Kohn-Rostocker Green's function formalism as implemented in the SPRKKR package \cite{sprkkr}. The shape of the potential was treated by the Atomic Sphere Approximation (ASA), while the relativistic effects were considered by taking the fully relativistic Dirac equation. The effect of the exchange correlation part of the energy was treated by considering the Generalized Gradient Approximation (GGA) as devised by Perdew, Burke and Ernzerhof \cite{PhysRevLett.77.3865}.  Substitutional disorder between sub-lattices can also be studied by making the use of CPA. The interatomic exchange interactions were calculated via the LKAG formalism. \cite{lkag}.

The UppASD code \cite{uppasd} is based on the framework of LLG formalism.
The temporal evolution of an atomic moment in LLG formalism is given by\cite{asd1,asd2},
\begin{equation}
\frac{d{\bf{m}_i}(t)}{dt}=-\frac{\gamma}{1+\alpha^2}{\bf{m}_i}(t)\times [{\bf {\cal H}^{i}_{eff}}+\frac{\alpha}{m_s}~({\bf{m}_i}(t)\times {\bf {\cal H}^{i}_{eff}})] \nonumber
\end{equation}
where ${\bf{m}_i}(t)$ is the atomic moment on the i$^{th}$ site at time t.
$\gamma$ is the gyromagnetic ratio and $\alpha$ is the Gilbert damping factor, which we have assumed to be 3$\times 10^{-4}$. $m_s$ is the
saturation moment for the i$^{th}$ atom. The temperature effect in ASD has been taken through a stochastic magnetic field included in the term of effective field ${\bf{\cal H}^{i}_{eff}}$.
The effective field ${\bf{\cal H}^{i}_{eff}}$ on the $i^{th}$ atom is calculated from the effective magnetic Hamiltonian given by
\begin{equation}
{\cal H}_{Mag}={\cal H}_{ex} \nonumber
\end{equation}
through $${\bf {\cal H}^{i}_{eff}}=-\frac{\partial {\cal H}_{Mag}}{\partial{\bf{m}}_i(t)}.$$
${\cal H}_{ex}$ describes the magnetic exchange interactions between the spins. ${\cal H}_{Mag}$ is given by
$$
{\cal H}_{Mag}=\sum_{i,j}J_{ij}{\bf S}_i\cdot{\bf S}_j
$$
In order to calculate magnon dispersion relation, one should calculate spin-spin correlation function. The spin-spin correlation in an effective field is obtained by solving the LLG equation and can be written as,
\begin{equation}
C({{ R}-{ R'}},t)= \langle m_{{ R}}(t) m_{{R'}} (0) \rangle -  \langle m_{{ R}}(t)\rangle \langle m_{{ R'}} (0) \rangle  \nonumber
\end{equation}
where the ensemble average is represented in the angular brackets and {\it k} is the cartesian component. The Fourier transform of the spin-spin correlation function, known as dynamical structure factor is written as,
\begin{equation}
S({\vec {q}} ,\omega) = \frac{1}{\sqrt{2\pi N}} \sum_{{ R}, { R'}} \int^{\infty}_{-\infty} e^{i \omega t} C({ R} - {R'}, t) dt \nonumber
\end{equation}
$S({\vec {q}},\omega)$ is measured in neutron scattering experiments.

\section{Results and Discussion}
\label{result}

\begin{figure*}[ht]
\centering
\includegraphics[scale=0.2]{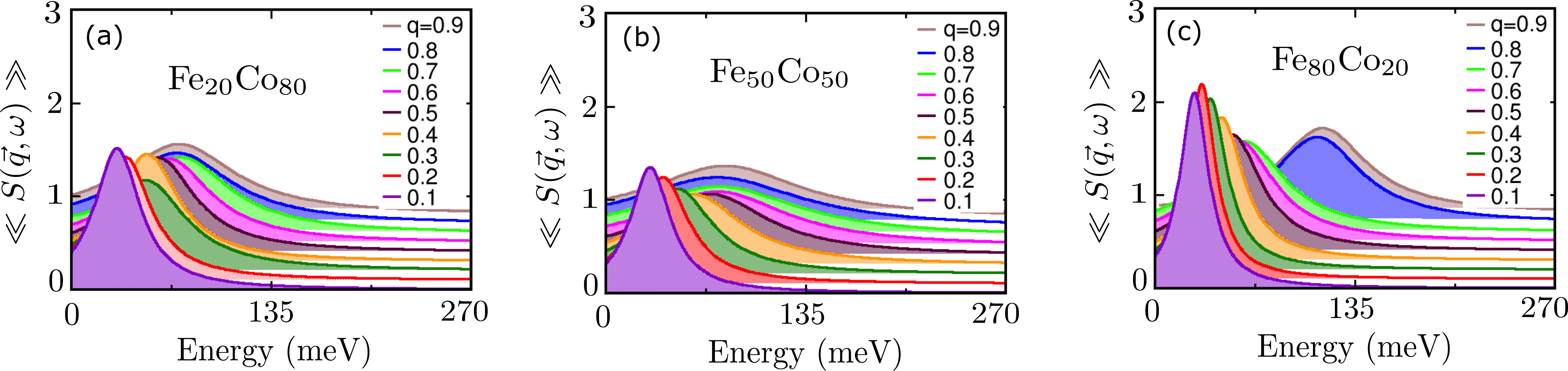}
\caption{The dynamical structure factors $\ll S(\vec{q},\omega)\gg$ for a selection of wave-vectors $\vec{{q}}$ varying from 0.1 to 0.9 along the symmetric $\Gamma-H$ direction in the Brillouin zone of bcc Fe$_x$Co$_{1-x}$ alloys with (a) x=0.2 (b) x=0.5 (c) x=0.8 respectively. In order to identify the peak position the dynamical structure function $\ll S(\vec{q},\omega)\gg$ is convoluted with a Lorentzian function, normalized to unity. 
\label{stuc}}
\end{figure*}

\begin{table}[ht]
 \small
\caption{The charge redistribution on alloying : showing both the distribution for the ordered
 B2 and disordered body centered cubic (BCC) structures for the 50-50 FeCo alloy. Charge is given in units of electronic charge $e$. }	\vskip 0.2cm
	\begin{tabular}{|c||cccccc|}\hline
Alloy&    \multicolumn{6}{|c|}{\bf Fe$_{0.5}$Co$_{0.5}$}\\ 
Component &\multicolumn{3}{c|}{\bf Fe}&\multicolumn{3}{|c|}{\bf Co} \\
\hline
Atomic Radius &\multicolumn{3}{c|}{\rule{0mm}{4mm} R$_0$=2.64{\rm\AA}}&\multicolumn{3}{|c|}{ R$_0$=2.60{\rm\AA}} \\
\hline
Charge &\multicolumn{1}{|c|}{\sl sp}& \multicolumn{1}{c}{\sl d}&\multicolumn{1}{|c|}{Tot}&
 \multicolumn{1}{|c|}{\sl sp}& \multicolumn{1}{c}{\sl d}&\multicolumn{1}{|c|}{Tot}
 \\ \hline
Atomic state &\multicolumn{1}{|c|}{2.0}& \multicolumn{1}{c}{6.0}&\multicolumn{1}{|c|}{8.0}&
 \multicolumn{1}{|c|}{2.0}& \multicolumn{1}{c}{7.0}&\multicolumn{1}{|c|}{9.0}
 \\ \hline
B2 ordered & \multicolumn{1}{|c|}{1.44}& \multicolumn{1}{c}{6.52}&\multicolumn{1}{|c|}{7.96}&
 \multicolumn{1}{|c|}{1.46}& \multicolumn{1}{c}{7.58}&\multicolumn{1}{|c|}{9.04}
 \\ \hline
BCC disordered &  \multicolumn{1}{|c|}{1.43}& \multicolumn{1}{c}{6.55}&\multicolumn{1}{|c|}{7.99}&
 \multicolumn{1}{|c|}{1.43}& \multicolumn{1}{c}{7.63}&\multicolumn{1}{|c|}{9.06}
 \\ \hline
 \end{tabular}
 \label{tab1}
\end{table}

\begin{figure*}[ht]
\centering
\includegraphics[scale=0.2]{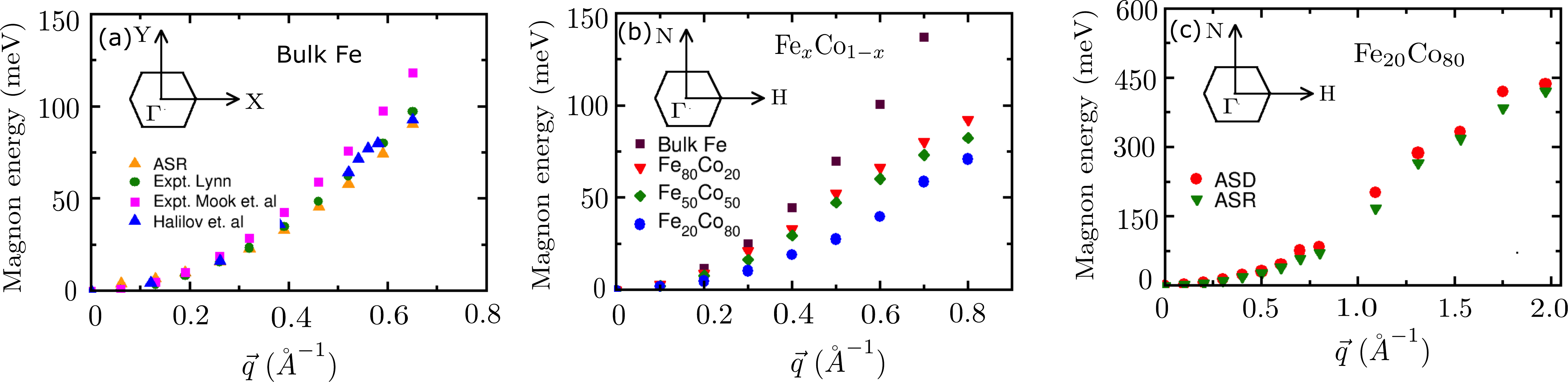}
\caption{(a)  Magnon dispersion
spectrum for bulk bcc Fe along the $\Gamma-Y$ direction from ASR. For comparison, the experimental spin-wave spectrum of bulk bcc Fe by Lynn [Ref \cite{lynn}] and another experimental data corresponding to a spin-wave stiffness constant of D = 280 ${\AA}^{2}$ by Mook {\etal} [Ref \cite{mook}] are also displayed. The blue triangles the present theoretical data by Halilov {\etal} [Ref \cite{halilov}]. All comparison are matches well with the data from ASR.  \cite{mook}. (b) Magnon dispersion spectrum for bulk, bcc Fe$_x$Co$_{1-x}$ alloys along the high symmetric $\Gamma-H$ direction in the Brillouin zone for three different Co concentration from ASR.  This shows a magnon softening in Fe$_{20}$Co$_{80}$ alloy compared to others.
(c) Magnon dispersion spectra for disordered Fe$_{20}$Co$_{80}$ from ASR and ASD.  
 \label{disp} }
\end{figure*}

\par Before studying the magnetization dynamics in Fe$_x$Co$_{1-x}$, we investigate the charge distribution of body-centered, disordered Fe$_{50}$Co$_{50}$ alloy. We compare the disordered alloy with the corresponding ordered, B2 structured alloy. In Table \ref{tab1}, we show the chemical effects of disorder and the redistribution of charge on alloying.  In these alloys, there is a small decrease of $sp$ like charges and a corresponding small increase in the $d$ like channel in both Fe and Co. This effects are small with the introduction of disorder.  The magnetocrystalline anisotropy is the energy difference between the magnetic ground state when the magnetic quantization (spin) axis is along easy axis (which is along $c$-axis for FeCo) and along easy plane (which is $ab$-plane for FeCo) i.e K$^u$ = E$^c$ - E$^{||}$. Bulk FeCo has $\frac{c}{a} \sim 1$ without any tetragonal distortion.  Therefore we neglect here the effect of single-ion anisotropy of FeCo alloys.

\par We are interested to calculate the 
ensemble averaged dynamical structure factor $\ll  S(\vec{q},\omega)\gg $ to understand the 
magnetization dynamics in disordered Fe$_x$Co$_{1-x}$ alloys. Figure \ref{stuc} (a)-(c) show the momentum resolved dynamical structure function $\ll  S(\vec{q},\omega)\gg$ convoluted with a Lorenzian function with three disorder concentrations ($x$ = 0.2, 0.5, 0.8) along the symmetric $\Gamma-H$ direction. Irrespective of the disorder concentration, for low $\vec{{q}}$ values, the spectral intensity, estimated from the peak value of the Lorenztian, is high. The spectral intensity drops to minimum values for intermediate $\vec{{q}}$, after that it again rises  with increasing $\vec{{q}}$. At the same time, the peak width monotonically increases with increasing energy as well as wave-vector ($\vec{{q}}$). The broadening of peak in $\ll  S(\vec{q},\omega)\gg $ for larger $\vec{{q}}$ values comes from magnon-electron scattering mechanism due to chemical impurity. The itinerant collective excitations of spin waves (magnon) lose their energy as a consequence of chemical disorder. A large damping of magnons in the tetragonally distorted bulk FeCo compound is also predicted previously \cite{sasi}, \cite{Jako}.

\begin{figure*}[ht]
\begin{center}
\includegraphics[scale=0.32]{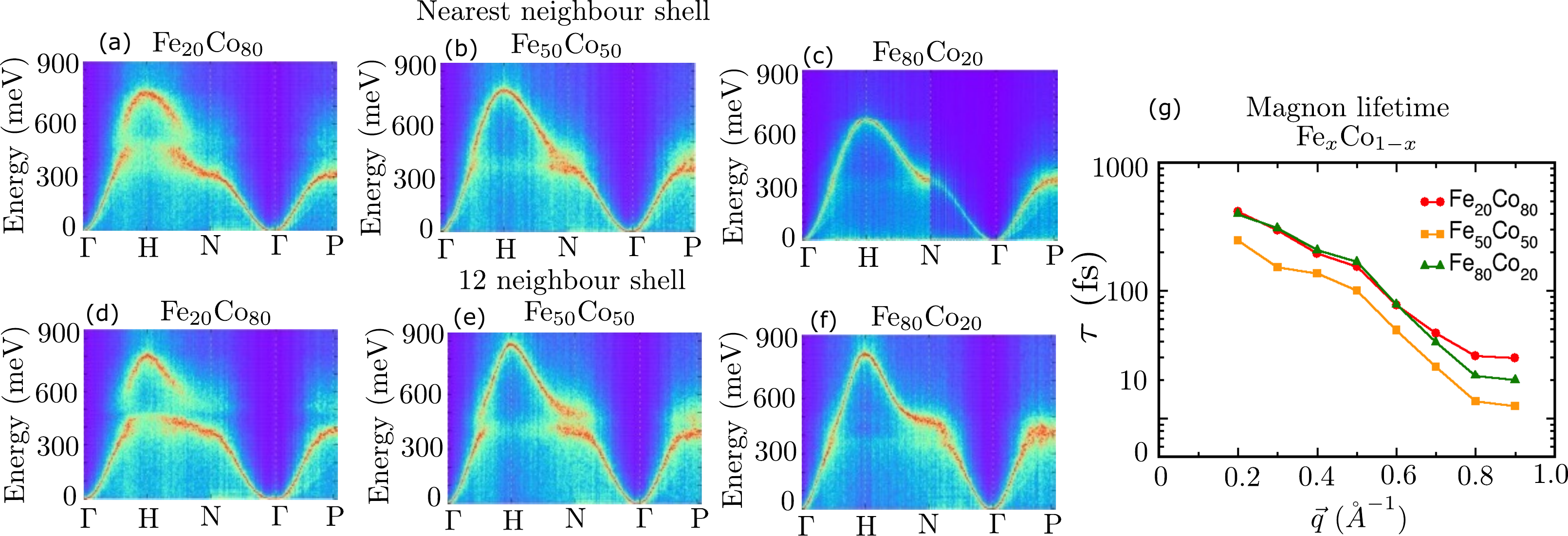}
\end{center}
\caption{The upper row figures ((a)-(c)) show the magnon dispersion of Fe$_{x}$Co$_{1-x}$ (where x = 0.2, 0.5 and 0.8) alloys with nearest neighbour exchange parameters and the lower row has the plots ((d)-(f)) with 12 neighbour shell's exchange parameters. (g) Magnon lifetime for disordered Fe$_{x}$Co$_{1-x}$ alloys for x= 0.2, 0.5 and 0.8 from ASR.
\label{asd_disp}}
\end{figure*}

 \begin{table}[ht]
    \small
    \caption{The calculated nearest-neighbour exchange parameters in the framework of ab-initio density functional theory using the Korringa-Kohn-Rostocker Green’s function formalism as implemented in the SPRKKR package \cite{sprkkr}. }\vskip 0.2cm
    \begin{tabular*}{0.45\textwidth}{ p{1.8cm} p{2.0 cm} p {2.0 cm} p{2.0 cm} }
    \hline\hline
        Alloy  & $Fe-Fe$  & $Fe-Co$  & $Co-Co$  \\
          & (meV)  & (meV)  & (meV)  \\
    \hline
     Fe$_{80}$Co$_{20}$ & 2.065 & 2.302 & 1.740  \\
     Fe$_{50}$Co$_{50}$ & 2.083 & 2.117 & 1.418 \\
     Fe$_{20}$Co$_{80}$  & 1.919 & 1.880 & 1.208 \\
    \hline
     \end{tabular*}
 \label{tab2}
\end{table}

\par Now, we shall elaborate the technique of peak finding from $\ll  S(\vec{q},\omega)\gg $. 
After configuration averaging, in order to  identify the position  of the intensity peaks of dynamical structure factor $\ll S(\vec{q},\omega) \gg$, we convolute the respective intensity profile of $\ll  S(\vec{q},\omega)\gg $ for each $\vec{{q}}$ vector with a Lorentzian function, normalized to unity. In order to justify this technique, we apply it to calculate magnon dispersion of bulk bcc Fe (see Fig.\ref{disp} (a)). This refers to the fact that the implemented technique is able to predict the experimental finding qualitatively.

\par Having stated this fact, we shall now employ our technique for chemically disordered Fe$_x$Co$_{1-x}$ alloys. Here we calculated the magnon dispersion of disordered Fe$_{x}$Co$_{1-x}$ alloys along the $\Gamma-H$ direction and study the effects of chemical disorder on it. We find the quadratic dependence upon $\vec{{q}}$ for the spin wave spectrum of magnon as shown in Fig.\ref{disp} (b). Bulk Fe has a more steeper magnon dispersion compared to the disordered structure. This is expected due to the absence of magnon-electron scattering for the disorder free case. The interestig point to note here is that the steepness of magnon energy gradually reduces  with increasing Co concentration. Fe$_{80}$Co$_{20}$ exhibits higher magnon velocity for less scattering effect and vice versa.

\par The magnon softening is clearly visible in Fe$_{20}$Co$_{80}$ at the zone boundary. This is due to the fact that magnon-electron scattering modifies the exchange energy between the Fe and Co moments. The chemical disorder between the Fe and Co atoms influences the hybridization between Fe and Co states, which reduces the exchange interactions. This is shown in table \ref{tab2}. The strength of exchange interaction for Fe$_{x}$Co$_{1-x}$ gradually decreases with increasing Co content. It clearly indicates the softening of Fe$_{x}$Co$_{1-x}$ with increasing Co content. The chemical disorder between Fe and Co also increases the damping which enhances the magnon softening. Ebert {\etal} calculated Gilbert damping parameter from first principles theory for bcc Fe$_{x}$Co$_{1-x}$ alloys  \cite{ebert}. Damping is minimum for 20\% Co content and gradually increases with increasing Co content. This leads to strong magnon softening in Fe$_{20}$Co$_{80}$ alloy. Our result is also consistent with the experiment by Oogane {\etal} on bcc Fe$_{x}$Co$_{1-x}$ alloy  \cite{oogane}. 

\par For completeness, we also compute the magnon dispersion spectrum along $\Gamma-H$ direction in the Brillouin zone. We compare the magnon spectra of disordered Fe$_{20}$Co$_{80}$ between ASR and ASD taking first nearest neighbour exchange interaction into consideration (as shown in Fig.\ref{disp} (c)). The ASD simulations are performed with a system size of 120$\times$120$\times$120 at a temperature of 0.1K. ASD follows similar characteristic behavior with increasing Co concentration. It is observed that the softening becomes stronger near the zone boundary (higher $\vec{q}$ values). The steepness of dispersion reduces as the magnon-electron scattering increases. It is also observed in the profile of $\ll  S(\vec{q},\omega)\gg $ in Fig.\ref{stuc} (a). Both the numerical techniques exhibit same effect.

\par To justify our results further, we perform  ASD simulations for nearest neighbour (Fig. \ref{asd_disp}(a)-(c)) and 12 neighbor exchange interactions (Fig.\ref{asd_disp} (d)-()f)). The qualitative features in the magnon spectra for first and 12 neighbor shells are similar. The softening of magnon spectra for Fe$_{20}$Co$_{80}$ alloy is similar to that obtained from ASR calculations shown above. Additionally, we observe some branching in dispersion 
along $H-N$ for higher Co concentration. A stronger tendency for branching is visible at the N point of the Brillouin zone.  However, here we focus only on the magnon softening in $\Gamma-H$ direction.  This strong magnon softening is in contradiction with previous study based on an itinerant electron model at $T = 0$ K  \cite{itin1,  itin2}. This revealed the possibility that a disordered Fe$_{x}$Co$_{1-x}$ may not be a simple itinerant ferromagnet. The effects of spin correlations are important for this system. This behaviour is commonly interpreted as a transition from conductivity like behaviour. This reflects the dominance of intra and interband transition. It is related to the broadening of electron energy bands caused by the increase of scattering events with concentration of Co. Increase of the Co content in Fe$_{x}$Co$_{1-x}$ leads to more impurity-scattering. This is responsible for band broadening which increases the damping of magnon. This effect completely suppresses the conductivity like behaviour in the low temperature regime and increases the scattering due to chemical disorder with increasing Co content.

\par  The broadening of the magnon excitation peak ($\ll  S(\vec{q},\omega)\gg $) provides a way to calculate the magnon lifetime $\tau $. The fourier transform of the Lorentzian in energy (or frequency) domain is an exponential decay of magnon lifetime like $\exp(-t\Gamma/2\hbar)$, where $\Gamma $ represents the intrinsic linewidth of the Lorentzian peak in energy and $\hbar$ is the reduced Plank constant. The relaxing magnetic modes or patterns are labelled by $\vec{{q}}$, such that the average `size' of the mode is $~O({q}^{-1})$. We have obtained the magnon lifetimes from the fourier transform of the configuration averaged correlation function. The large broadening of $\ll  S(\vec{q},\omega)\gg $  yields a small relaxation time which clearly indicates strong scattering. Fig.\ref{asd_disp}(g) shows the lifetime $\tau$ for Fe$_{x}$Co$_{1-x}$ (x=0.8, 0.5, 0.2) as a function of $\vec{{q}}$ (${\AA}^{-1}$).  Our study reveals the minimum magnon lifetime in Fe$_{50}$Co$_{50}$. It  ensures that the disorder is maximum in 50-50 configuration where $\ll  S(\vec{q},\omega)\gg $ becomes maximally broadened for higher $\vec{q}$ values (see Fig.\ref{stuc} (b). Therefore, the magnon lifetime is not governed by hybridization  between Fe and Co rather controlled by disorder configuration.

\par  In this current study,  we propose the ASR technique to study magnatization dynamics in chemically disordered alloys (which could also be generalised for surfaces and interfaces) and compare calculated magnon energy spectra with ASD simulated magnon spectra.  There are two different approaches within first-principles theory for studying magnatization dynamics in materials, one is ASR and other is ASD which are based on two different working principles.  ASD is based on the LLG theory, whereas the ASR based on Kubo linear response theory.  ASR captures the effect of hybridization between Fe and Co on the magnon softening of Fe$_{20}$Co$_{80}$ alloy (see Fig.\ref{disp}(c)) as well as magnon life time in chemically disordered alloys originating from magnon band broadening (see Fig.\ref{asd_disp}(g)).

\section{Conclusion}
\label{conclusion} 

\par In conclusion, we have investigated magnetization dynamics in presence of chemical disorder using first-principles TB-LMTO-ASR method interfaced with Kubo linear response theory. This serves as a  general and powerful computational tool for studying the effect of chemical disorder on magnetic excitations in binary random alloys. We apply our technique to study the magnetization dynamics in disordered Fe$_{x}$Co$_{1-x}$ alloys, which are of great fundamental as well as technological interest  \cite{zak,  ike,  liu1, liu2}. We show that magnon-electron scattering mechanism plays a crucial role in determining the exchange interaction between Fe and Co. The magnon softening increases with Co concentration as the exchange interaction decreases. We verified this finding in two high symmetry directions in the Brillouin zone. Our findings are further supported  by atomistic spin dynamics simulations. Moreover, we calculate magnon lifetimes from the broadening widths of dynamical structure factors. It is found that the maximally disordered alloy shows the minimum lifetime. Therefore, the magnon softening depends on the hybridization between Fe and Co while magnon lifetime depends on the disorder configuration. The proposed formalism can be used to study the magnetization dynamics in any binary random alloy.



%


\end{document}